\documentclass[10pt,twocolumn]{paper}
\usepackage{epsfig, graphicx}
\usepackage[english]{babel}
\usepackage[margin=1.5cm]{geometry}

\title{\center \rm \bf  X-Ray Study of Thermotropic Mesophases of
an n-Triacontanol Adsorption Film \\ at the n-Hexadecane -- Water Interface}

\author{\small \rm Aleksey M. Tikhonov\/\thanks{tikhonov@kapitza.ras.ru}\\ \small
Kapitza Institute for Physical Problems, Russian Academy of Sciences, Moscow, Russia}

\begin{document}
\maketitle

\abstract{ \it  Using synchrotron radiation with a photon energy of 15 keV, the molecular structure of an adsorbed n-triacontanol layer at the n-hexadecane -- water interface in different its phase states has been studied by the method of diffuse X-ray scattering. The analysis of the experimental data shows that a transition to the multilayer adsorption occurs at a temperature below the two-dimensional vapor -- liquid transition at the interface. This transition has been attributed to a feature in the temperature dependence of the concentration of micelles in a surface layer $\sim 100\div 200$\,\AA{} thick.  }

\vspace{0.25in}

Various reversible two-dimensional transitions between surface mesophases of fatty alcohols and
acids (lipids) are possible in an adsorbed film at the n-alkane - water interface [1-7]. In particular, a solid -- vapor phase transition is observed in a fluoroalkanol Gibbs monolayer [5, 8] and a solid -- liquid phase transition is observed in a carbon acid monolayer [9]. It was briefly reported in [10] that a liquid -- vapor thermotropic phase transition occurs in adsorbed n-triacontanol (C$_{30}$-alcohol) film at the n-hexadecane -- water interface. In this work, the molecular structure of the neutral surface mesophases of this lipid is studied by diffuse (nonspecular) X-ray scattering and reflectometry with the use of synchrotron radiation. It is shown that a transition from the structure with a width of $\sim 3$\,\AA{} to a monolayer with a thickness of $\approx 27$\,\AA{} and, then, to a structure $100\div 200$\,\AA{} thick occurs with decreasing temperature in a relatively narrow temperature range. We attribute the latter transition to an increase in the concentration of micelles in the surface layer.

An adsorption film at the planar oil–water interface can be considered as a two-dimensional thermodynamic
system with the parameters ($p,\,T,\,c$), where $p$ is the pressure and $c$ is the concentration of the lipid in the volume of the hydrocarbon solvent [11-13]. According to [10], the liquid–vapor transition in the adsorbed C$_{30}$-alcohol film at the n-hexadecane -- water interface at $p=1$\,atm and $c\approx0.6$\,mmol/kg is observed at $T_{c}\approx 300$\,K. The corresponding temperature dependence of the interfacial tension $\gamma(T)$, measured by the Wilhelmy plate method, is shown by
closed circles in Fig. 1 [10, 14]. A change in the slope of is due to a change in the surface enthalpy at the transition $\Delta H = - T_c\Delta(\partial \gamma/\partial T)_{p,c}$ $=0.42\pm 0.04$\,J/m$^2$. At the same time, the transition to the C$_{30}$-alcohol monolayer at the n-hexane -- water interface (open circles in Fig. 1) is characterized by the tripled value $\Delta H =  1.3\pm 0.1$\,J/m$^2$.

The reflection coefficient $R$ and the intensity of diffuse surface scattering $I_n$ of X rays at the n-hexadecane -- water interface were measured at the X19C beamline of the National Synchrotron Light Source (NSLS, Brookhaven National Laboratory, United States) with the use of radiation with the wavelength
$\lambda=0.825 \pm 0.002$\,\AA{} [15].

Let {\bf k}$_{\rm in}$ and {\bf k}$_{\rm sc}$ be the wave vectors of the incident and scattered beams, respectively, with the amplitude $k_0= 2\pi/\lambda$ (see the inset of Fig. 2). In the coordinate system where the origin $O$ lies at the center of the illuminated region, the plane $xy$ coincides with the interface, the $Ox$ axis is perpendicular to the beam direction, and the axis $Oz$ is normal to the surface and is directed opposite to the gravitational force, the components of the scattering vector {\bf q = k$_{\rm in}$ {\rm -} k$_{\rm sc}$} in the interface plane are $q_x \approx k_0\phi$ and $q_y \approx k_0(\alpha^2-\beta^2)/2$ and the normal component is $q_z\approx k_0(\alpha+\beta)$ ($\alpha,\beta << 1$, $\phi \approx 0$).

\begin{figure}
\hspace{0.15in}
\epsfig{file=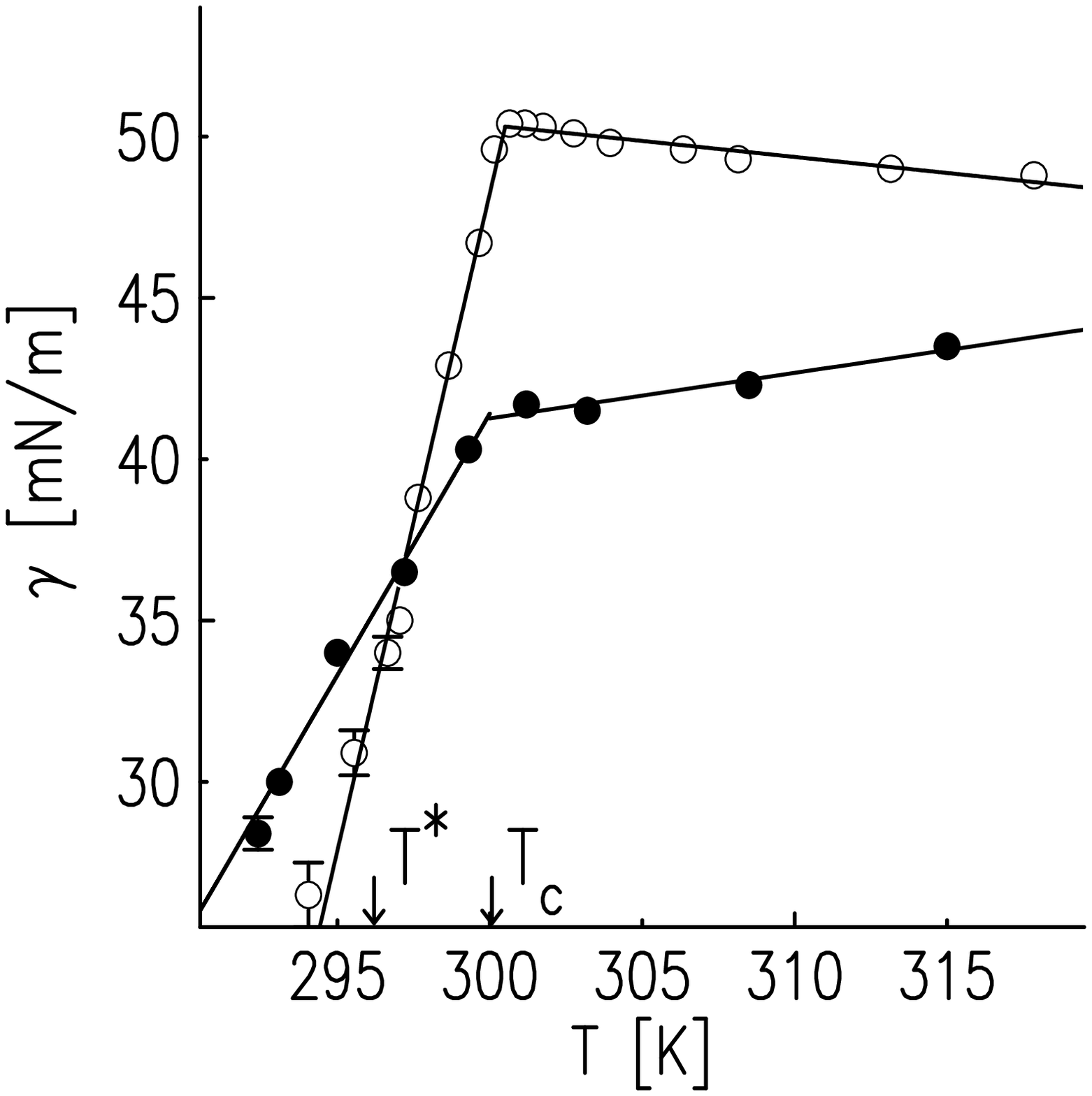, width=0.4\textwidth}

\small {\bf Figure 1.} \it Temperature dependences of the interfacial tension $\gamma(T)$ at  the n-hexane -- water (open circles) and  n-hexadecane -- water (closed circles) interfaces with the adsorbed n-triacontanol layer [10]. The straight lines are linear approximations of the rectilinear segments of $\gamma(T)$.
\end{figure}

\begin{figure}
\hspace{0.15in}
\epsfig{file=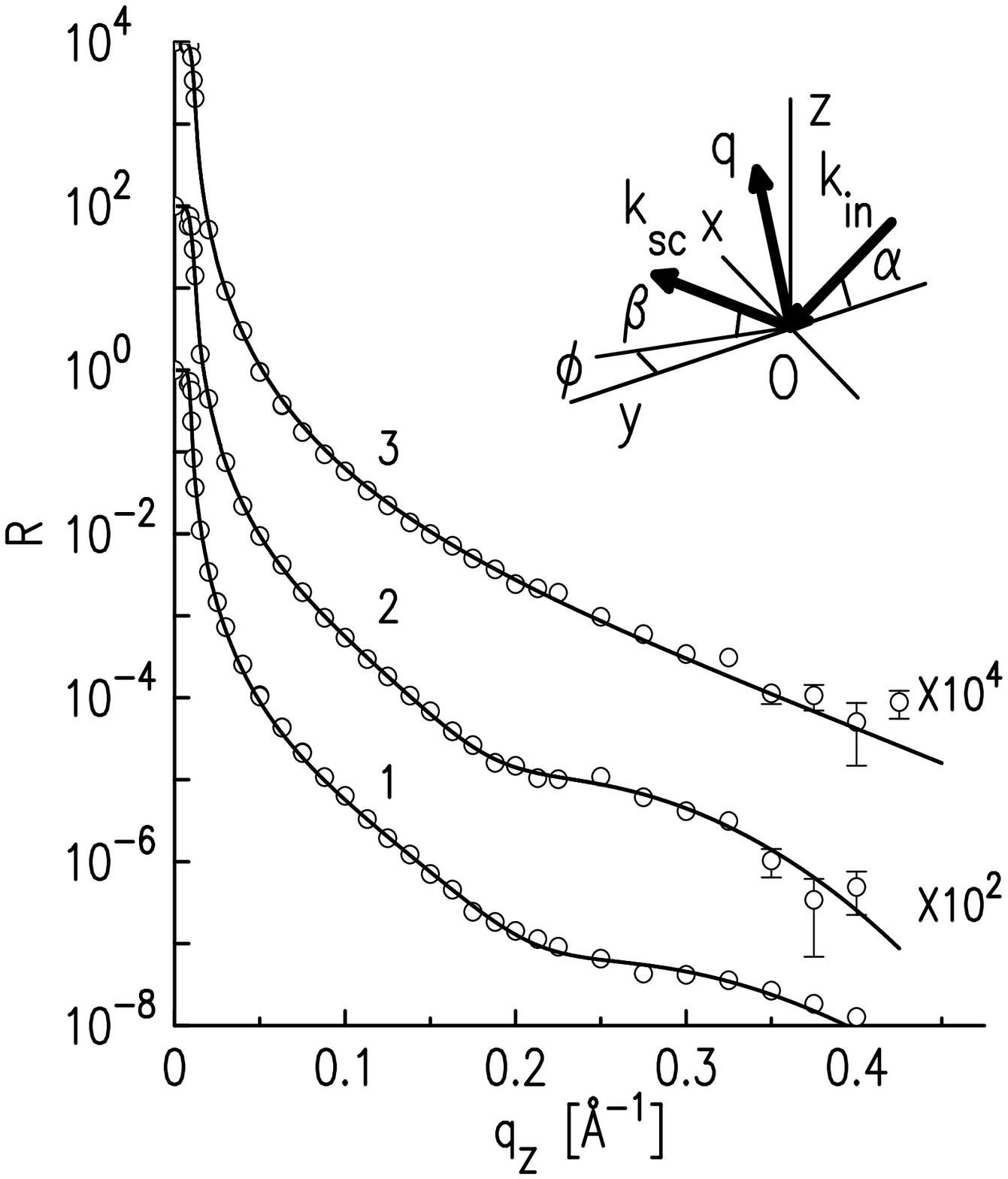, width=0.4\textwidth}

\small {\bf Figure 2.} \it Reflection coefficient $R$ versus $q_z$ for various phase states of the adsorbed n-triacontanol film at the n-hexadecane -- water interface: (1) multilayer at $296.1$\,K ($T<T^*$),
(2) liquid multilayer at 298.0\,K ($T^*<T<T_c$), and (3) vapor at 321.0\,K ($T>T_c$). The solid lines correspond to models of capillary wave structures. The inset shows the kinematics of surface scattering in the coordinate system where the $xy$ plane coincides with the n-hexadecane -- water interface, the $Ox$ axis is perpendicular to the beam direction, and the $Oz$ axis is perpendicular to the surface.

\end{figure}

According to the method described in [9, 16], the interface sample was prepared in a thermostatic cell,
which was then placed on an optical table with active vibration isolation. Deionized water (Barnstead,
NanoPureUV) with a volume of $\sim 100$\,mL was used as the bottom bulk phase. About $\sim 50$\,mL of the solution of n-triacontanol in n-hexadecane with $c\approx0.6$\,mmol/kg was used as the top bulk phase. Saturated hydrocarbon C$_{16}$H$_{34}$ (the melting temperature is 291\,K, the boiling temperature is 560\,K, and the density at 298\,K is $\approx 0.77$\,g/cm$^3$) was preliminarily purified by repetitive filtering in a chromatographic column [17]. Ñ$_{30}$-alcohol or C$_{30}$H$_{62}$O was doubly purified by recrystallization from the supersaturated solution in n-hexane.

Measurements of $R(q_z)$ at low $q_z$ values impose constraints on the longitudinal (along the $Oy$ axis)
dimension of the sample, which is $75$\,mm (the transverse dimension is 150\,mm). First, this is due to the
effect of boundary conditions near the walls of the cell on the planarity of the interface. Second, the longitudinal dimension of the illuminated region of the sample at the smallest glancing angle $\approx 4\cdot 10^{-4}$\,rad ( $q_z\approx 0.007$\,\AA$^{-1}$) and the smallest vertical dimension of the beam $\approx 10$\,$\mu$m is $\sim 30$\,mm. A sufficiently flat region of the n-hexadecane -- water interface with such a width applicable for the measurement of scattering was obtained only in cells thicker than $75$\,mm. 

The parameters of the optical measurement scheme were considered in detail in [9, 15, 18]. At small glancing angles, the vertical dimension of the incident beam is determined by slits spaced from the center of the cell by a distance of $\sim 120$\,mm and the natural divergence of the beam $\sim 10^{-4}$\,rad is reduced to $\sim 2\cdot 10^{-5}$\,rad by two input slits with a gap of $\sim 10$\,$\mu$m at a distance of $\sim 600$\,mm. In the region of large glancing angles ($q_z>0.2$\,\AA$^{-1}$), the maximum vertical dimension of the input slits of 0.4\,mm at measurements of is specified by the chosen vertical angular resolution of the detector in the $yz$ plane, $\Delta\beta\approx 1.2\cdot10^{-3}$\,rad (the slit with a vertical gap of 0.8\,mm at a distance of $\approx680$\,mm from the center of the sample).

Figure 2 shows the dependences $R(q_z)$ for various phase states of the adsorbed n-triacontanol film. At
$q_z < q_c=(4\pi/\lambda)\alpha_c$$\approx 0.01$\,\AA$^{-1}$, the incident beam undergoes total external reflection; i.e., $R\approx 1$. The total external reflection angle $\alpha_c=\lambda\sqrt{r_e\Delta\rho/\pi}$ $\approx 6\cdot10^{-4}$\,rad (where $r_e=2.814\cdot10^{-5}$\,\AA{} is the classical electron radius) for the n-hexadecane -- water interface is determined by the difference $\Delta\rho=\rho_w-\rho_{h}$ between the bulk electron densities in the hydrocarbon solvent $\rho_{h}$$\approx0.27$\,{\it e$^-$/}{\AA}$^3$ and in water $\rho_w\approx0.33$\,{\it e$^-$/}{\AA}$^3$.

Figure 3 shows data for the normalized intensity of diffuse surface scattering $I_n (\beta) \equiv (I(\beta)-I_b(\beta))/I_0$ (the normalization condition is $I_n(\alpha)\equiv 1$) obtained at the glancing angle $\alpha \approx 3.3 \cdot 10^{-3}$\,rad ($\approx 0.19^\circ$) for various phase states of the interface. Here, $I(\beta)$ is the number of photons scattered by the bulk of the sample and reflected (specularly and diffusely) from the surface in the illuminated region with an area of $A_0\approx 30$\,mm$^2$ at the center of the interface in the $\beta$ direction; $I_0$ is the normalization constant proportional to the intensity of the incident beam, which was controlled in the experiment immediately before entry of the beam into the cell; and $I_b(\beta)$ is the number of photons scattered in the bulk of n-hexadecane on the path to the interface, which is determined by the method described in detail in [16]. The most intense peak on the curve $I_n(\beta)$ corresponds to specular reflection at $\beta = \alpha$, and the peak against the diffuse background at $\beta \to 0$ illustrates an increase in the scattering intensity at $\beta = \alpha_c$ [19]. The measurement of $I_n(\beta)$ was performed with a collimated beam with the angular divergence in the vertical plane $\Delta\alpha \approx  5\cdot10^{-5}$\,rad and $\Delta\beta \approx 3\cdot10^{-4}$\,rad.

\begin{figure}
\hspace{0.15in}
\epsfig{file=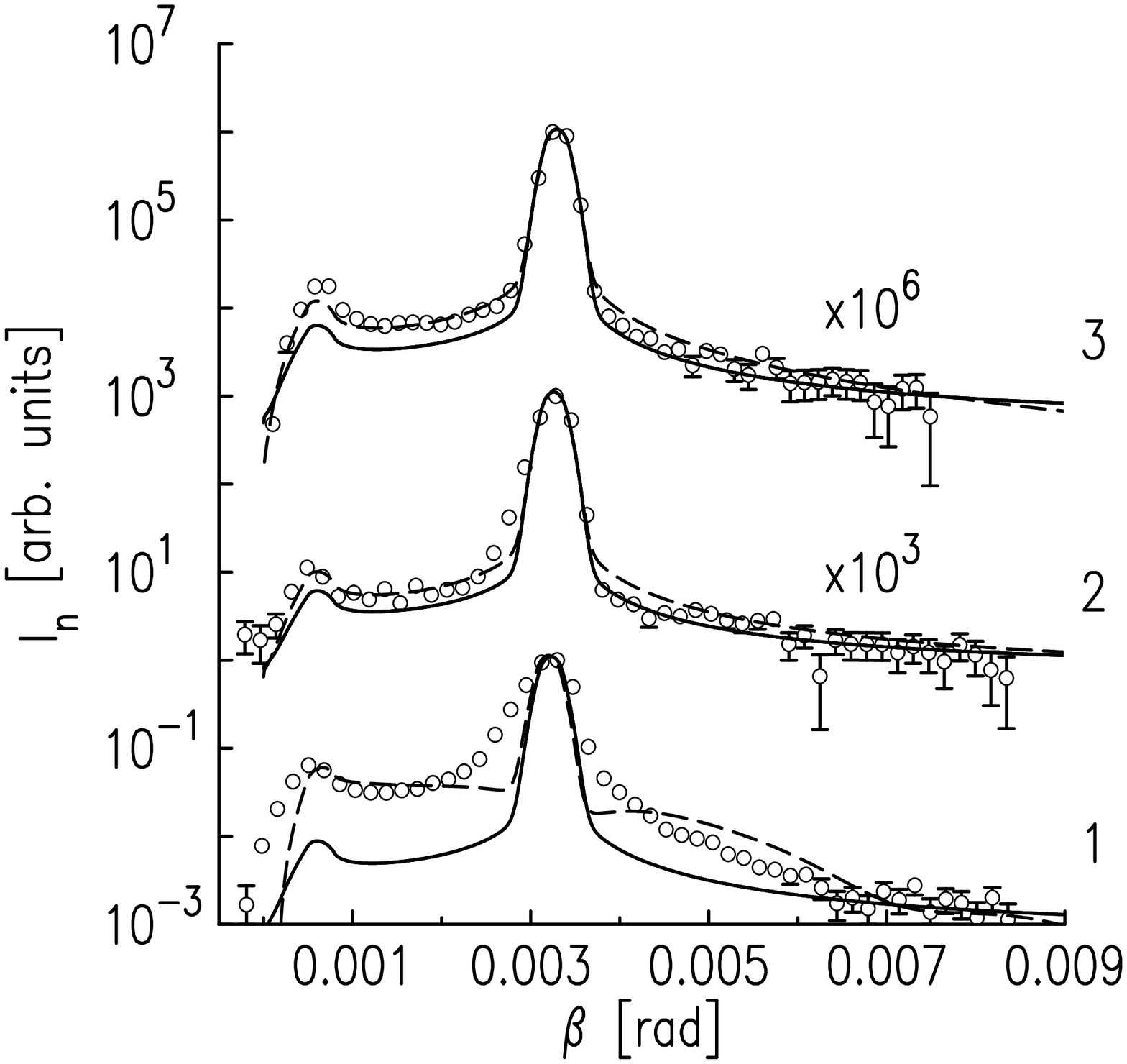, width=0.4\textwidth}

\small {\bf Figure 3.} \it Angular dependencies of the surface scattering intensity $I_n$ at the glancing angle $\alpha \approx 3.3 \cdot 10^{-3}$\,rad for various phase states of the adsorbed n-triacontanol film at the n-hexadecane -- water interface: (1) 296.0\,K (multilayer, $T<T^*$); (2) 298.0\,K (liquid monolayer, $T^*<T<T_c$); and (3) 325.2\,K (vapor, $T>T_c$). The solid and dashed lines correspond to the models of capillary and noncapillary wave structures, respectively.

\end{figure}

From data for $R(q_z)$ and $I_n (\beta)$, we obtain information on the surface normal structure of the interface using the distorted wave Born approximation [20]. According to the model approach described in [16, 18], the interpretation of experimental data is reduced to determining the parameters of the structure factor function of the interface $\Phi(q)$, which is in turn specified by the chosen model of the electron density distribution $\langle \rho(z) \rangle$ across the interface. Symmetric model profiles $\langle \rho(z) \rangle$ are constructed with the error function ${\rm erf}(x)$, which is used in the standard theory of capillary waves [21].

\begin{figure}
\hspace{0.5in}
\epsfig{file=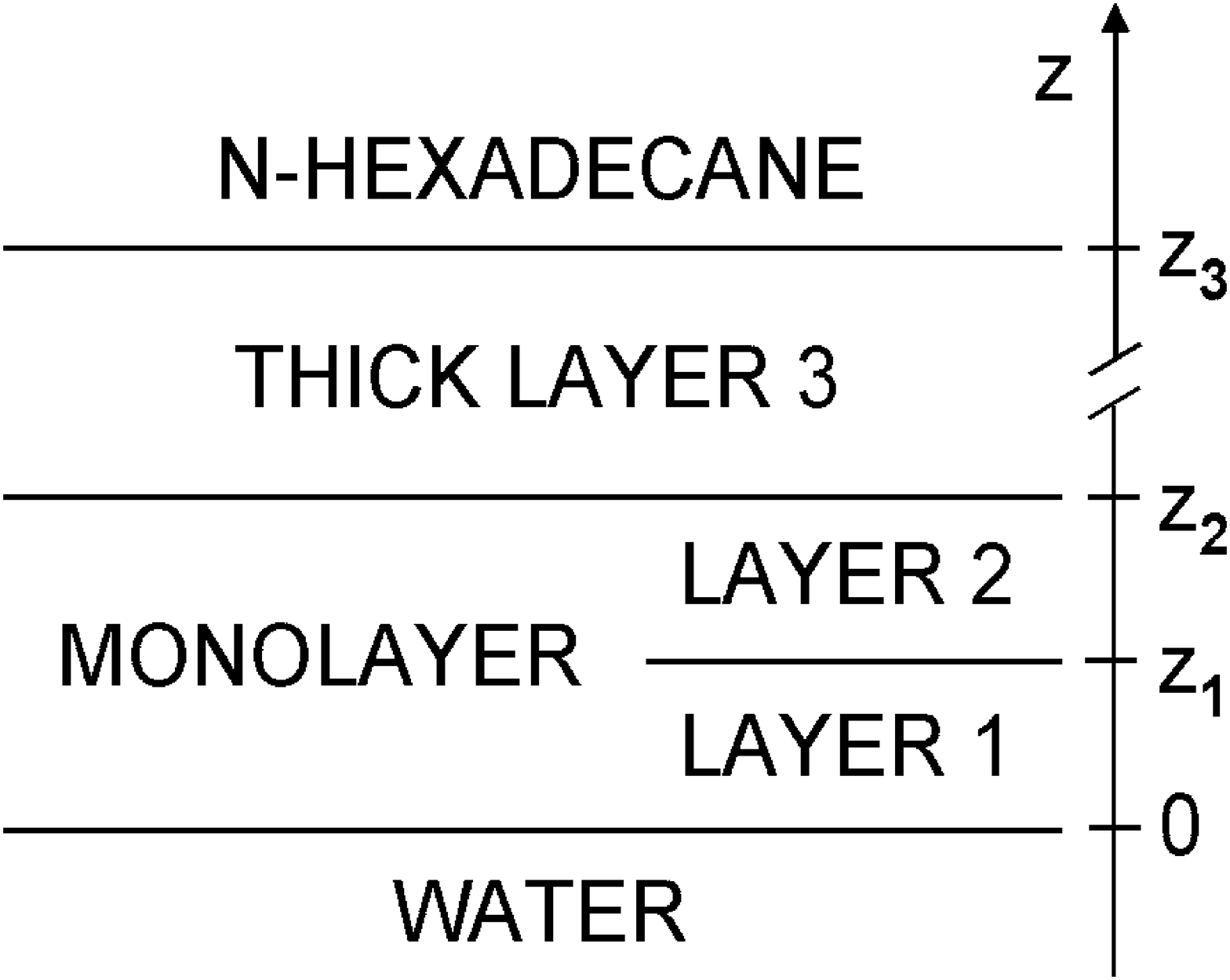, width=0.31\textwidth}

\small {\bf Figure 4.} \it Model of the transverse structure of adsorbed n-triacontanol C$_{30}$H$_{62}$O film at the n-hexadecane -- water interface.

\end{figure}

The qualitative model of the structure of the adsorbed C$_{30}$-alcohol film at the n-hexadecane -- water
interface shown in Fig. 4 provides a self-consistent interpretation of reflectometry and diffuse scattering
data with a minimum number of fitting parameters. Layers 1 and 2 describe the structure of the Gibbs
monolayer and are formed by polar head parts -CH$_2$OH (with a length of $\approx 2.4$\,\AA{}) and hydrophobic hydrocarbon tails -C$_{29}$H$_{59}$ (with a length of $\approx 38.3$\,\AA{}), respectively. As the temperature increases near $T_{c}$, a significant fraction of C$_{30}$H$_{62}$O molecules adsorbed in the Gibbs monolayer are evaporated from the interface and are dissolved in the bulk of the hydrocarbon solvent. Thus, the gas phase of the monolayer is implemented. Additional thick layer 3 is necessary to qualitatively explain a high intensity of grazing diffuse scattering, which exceeds the calculated value for the capillary wave channel of elastic scattering for all phase states of the adsorbed film.

At $T > T_c$, the dependences $R(q_z)$ and $I_n(\beta)$ in the gas phase of the interface are fairly well described within a single-parameter model with the structure factor
\begin{equation}
\Phi(q)_V = e^{-\sigma^2 q^2/2}.
\end{equation}
The minimum value of the parameter $\sigma^2$, which determines the width squared of the interface, is limited by the "capillary width" squared:
\begin{equation}
\sigma_{0}^2 =  \frac{k_BT}{2\pi\gamma(T)} \ln\left(\frac{Q_{max}}{Q_{min}}\right),
\end{equation}
which is in turn specified by the short-wavelength limit in the spectrum of capillary waves
$Q_{max} = 2\pi/a$ (where $a\approx 10$ {\AA} is the intermolecular distance) and $Q_{min}=q_z^{max}\Delta\beta/2$ (where $q_z^{max}$ is the maximum $q_z$ value in the experiment) [22-26].

The dependence $R(q_z)$ calculated by Eq. (1) for $T>T_c$ with the fitting parameter $\sigma = 3.4\pm 0.2$\,\AA{} is shown by line 3 in Fig. 2. Within the error, $\sigma$ coincides with $\sigma_0 = (3.59 \pm 0.04)$\,\AA{} for this measurement. On one hand, this calculation without free parameters
describes the dependence $R(q_z)$. On the other hand, the observed diffuse scattering intensity at $T>T_c$ is noticeably higher than that calculated by Eqs. (1) and (2) (solid line 3 in Fig. 3). To describe $I_n(\beta)$ by means of Eq. (1), the fitting value $\sigma \approx 6.5$\,\AA{} should be taken for the effective width (dashed line 3 in Fig. 3); Eq. (2) gives $\sigma_0 \approx 4.3$\,\AA{} taking into account the difference in $\Delta\beta$ and $q_z^{max}$ in measurements of $I_n$ and $R$. This indicates that the interface has an internal structure with a width larger than $\sqrt{\sigma^2-\sigma_0^2}\sim 5$\,\AA, which has a noncapillary wave nature [27].

At $T < T_c$, the reflection coefficient $R(q_z)$ in the liquid phase of the Gibbs monolayer is fairly well
described within the qualitative two-layer model of the structure factor (lines 1 and 2 in Fig. 2):
\begin{equation}
\Phi(q)_{L} = \frac{e^{-\sigma^2q^2/2}}{\Delta\rho}\sum_{j=0}^{2}{(\rho_{j+1}-\rho_j) e^{-iq_zz_j}},
\end{equation}
where $z_0=0$, $\rho_0=\rho_w$, and $\rho_3 = \rho_h$. The relative electron densities are $\rho_1/ \rho_w = 1.10 \pm 0.03$ and $\rho_2/\rho_w = 0.90 \pm 0.03$ and the coordinates of the layer boundaries
are $z_1 = 10\pm 2$\,\AA{} and $z_2 \approx 27$\,\AA{}. The total thickness of the Gibbs monolayer is $z_2-z_0=(27 \pm 2)$\,\AA. The calculated values $\sigma_0 = 3.7\pm 0.1$\,\AA{} and $\sigma_0 = 4.1\pm 0.1$\,\AA for lines 1 and 2, respectively, coincide within the experimental error with the respective fitting values $\sigma = 3.8\pm 0.2$\,\AA{} and $\sigma = 4.3\pm 0.2$\,\AA{}.

The observed scattering intensity $I_n(\beta)$ in the range $T_c > T > T^* \approx 296$\,K is insignificantly higher than the calculated value (solid line 2 in Fig. 3) and can be described by Eq. (3) with the effective width $\sigma \approx 5.9$\,\AA{} (dashed line 2 in Fig. 3), which is larger than $\sigma_0\approx 4.3$\,\AA obtained from Eq. (2). 

Finally, intensity $I_n(\beta)$ increases significantly at $T<T^*$(see experimental points 1 in Fig. 3). Fitting Eq. (3) to all these data gives the width $\sigma \approx 30$\,\AA{}, whereas $\sigma_0\approx 5.4$\,\AA.

The range of angles of observation of the diffuse background in scattering experiments is limited to
$\beta<0.006$\,rad or $q_z<q_z^* \sim 0.07$\,\AA$^{-1}$, whereas the maximum value $q_z^{max}$ in reflectometry experiments is about $0.4$\,\AA$^{-1}$. On one hand, the reflectometry data are quite well described by the parameter calculated by Eq. (2). On the other hand, the effective roughness of the surface according to diffuse scattering data is $> 6$\,\AA, which can reasonably be attributed to the existence of an extended near-surface structure (layer 3 in Fig. 4) thicker than $2\pi/q_z^{*}\approx 100$\,\AA. Then, a high grazingscattering intensity at $T<T^*\approx 296$\,K can qualitatively be explained within a three-layer model (multilayer adsorption) [16]:
\begin{equation}
\displaystyle
\Phi(q)^*_{L} +\frac{ \displaystyle \delta\rho e^{-\sigma_3^2q_z^2/2 }}{ \displaystyle \Delta\rho }
e^{-iq_z z_3}.
\end{equation}
Here, the second term describes the third layer with the thickness $z_3-z_2$ and density $\rho_h + \delta\rho$, the parameter $\sigma_3$ reflects the noncapillary wave structure of the boundary of layer 3 with the solvent, and $\Phi(q)^*_{L}$ is given by Eq. (3) with the substitution $\rho_3 = \rho_h + \delta \rho$.

The intensity $I_n(\beta)$ calculated by Eq. (4) is shown by dashed line 1 in Fig. 3. The estimated thickness of the thick layer is $z_3-z_2 \approx 200$\,\AA, the parameter $\delta\rho/\rho_w$ is $0.02 - 0.09$, and the width is $\sigma_3 \approx 20 - 40$\,\AA. The density $\rho_h + \delta\rho$ corresponds to the electron density in a high-molecular-weight alkane liquid [28]. The experimentally observed broadening of the central peak on line 1 is possibly due to small-angle scattering from micelles in the bulk of n-hexadecane, which was disregarded in the calculations of $I_n(\beta)$.

Model profiles of the electron density $\langle \rho(z) \rangle$ for mesophases of the adsorbed C$_{30}$-alcohol film in units of $\rho_w$ are shown in Fig. 5. At $T > T_c$, the gas phase of the Gibbs monolayer (structure 3), which is characterized by a single parameter, the interface width $\sigma \approx 3.4$\,\AA{}, is implemented in the adsorbed film. In the range $T^* <T < T_c$, the liquid Gibbs monolayer with the thickness $(27\pm2)$\,\AA (structure 2) is implemented.
The observed diffuse scattering intensity in these phase states of the adsorbed film exceeds the calculated
value for the capillary wave channel of elastic scattering and indicates the presence of the weakly
contrast layer 3 with a thickness of $\sim 100$\,\AA in the surface structure. Structure 1 at $T< T^*$ differs from structure 2 in the presence of the dense ($\rho_3 \approx 0.9 \rho_w$) and thick ($\sim 200$\,\AA{}) layer 3. Such a structural change can be called multilayer adsorption.

We believe that the participation of n-triacontanol--micelle aggregates in the formation of the structure
of the adsorbed film can explain a surprisingly high background of diffuse scattering in all phase
states of the n-hexadecane -- water interface, which cannot be due to scattering on thermal fluctuations of
the interface. The characteristic diameter of a spherical micelle is about two lengths of the C$_{30}$-alcohol molecule, i.e., $\approx 80$\,\AA{}  ($\sim 2\pi/q_z^{*}$). The incomplete filling of surface layer 3 with a thickness of $\sim 200$\,\AA{} can be responsible for the observed blurring or a large width $\sigma_3 \approx 30$\,\AA{} of the interface between the adsorbed film and bulk.

\begin{figure}
\hspace{0.15in}
\epsfig{file=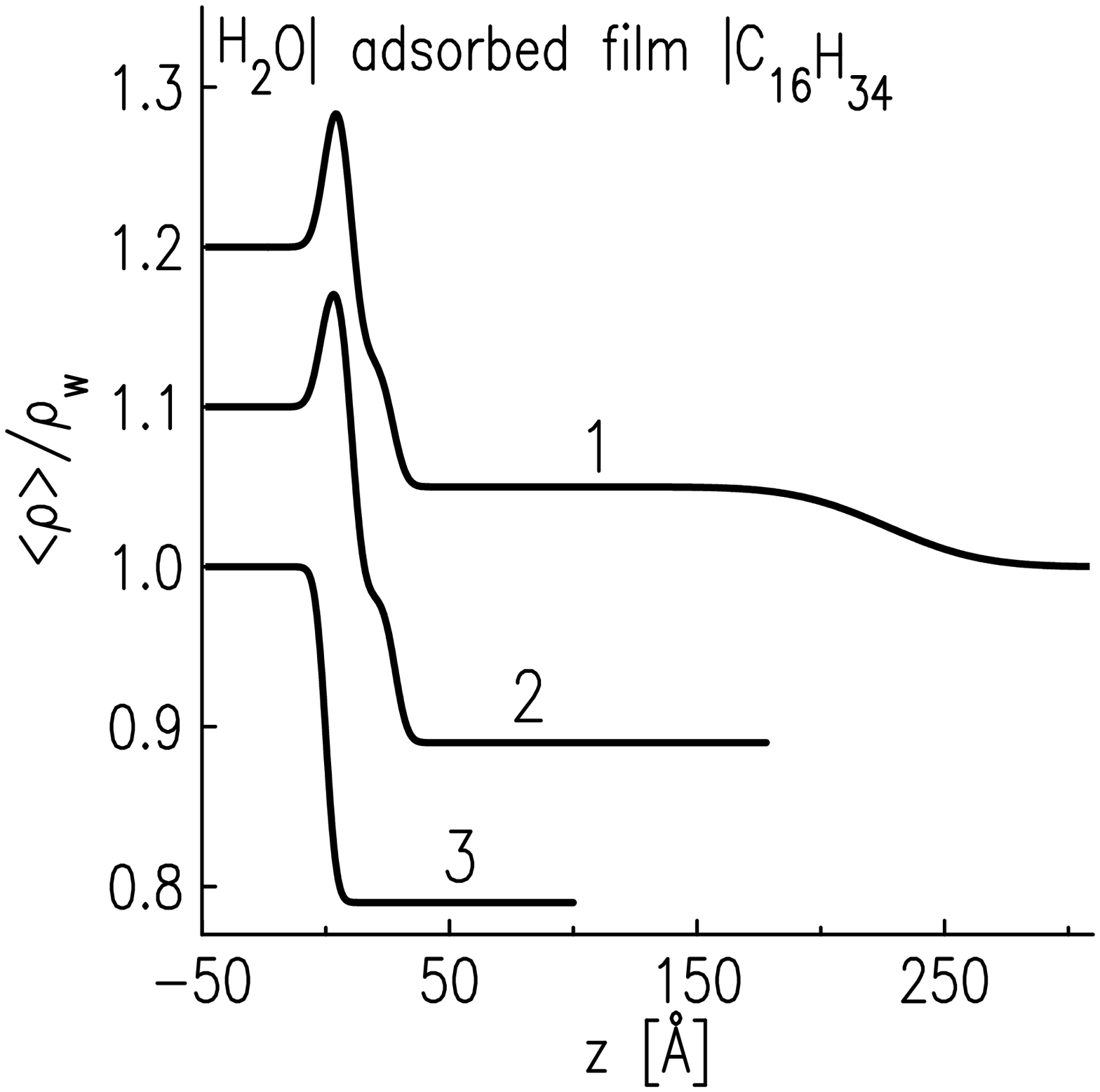, width=0.4\textwidth}

\small {\bf Figure 5.} \it Model profiles of the electron density $\langle \rho(z) \rangle$ of the adsorbed C$_{30}$-alcohol film in units of the electron density $\rho_w=0.333$ {\it e$^-$/}{\AA}$^3$ in water under normal conditions: (1) the model with an extended layer given by Eq. (4) at $T<T^*$, $\sigma=4.3$\,\AA{} è $\sigma_3=30$\,\AA{}; (2) the model of the Gibbs liquid monolayer given by Eq. (3) at $T^*<T<T_c$ and $\sigma= 3.8$\,\AA{}, and (3) the model of the gas phase given by Eq. (1) at $T>T_c$ and $\sigma= 3.4$\,\AA{}. For convenient comparison, profiles 2 and 1 are shifted along the $y$ axis by 0.1 and 0.2, respectively. The position of the interface between the polar region of n-triacontanol molecules and water is placed at $z = 0$.
\end{figure}

The described structures of neutral surface mesophases of the C$_{30}$-alcohol at the n-hexadecane -- water interface noticeably differ from the structure of both the solid phase of its Langmuir monolayer on the water surface and its mesophases at the n-hexane -- water interface [29]. In particular, the observed thickness $(27\pm 2)$\,\AA{} of the n-triacontanol Gibbs monolayer at the n-hexadecane -- water interface (area per molecule is $A = (29\pm 3)$\,\AA$^2${}) is noticeably smaller than $(36\pm 2)$\,\AA{} (area $A = (24 \pm 1$)\,\AA$^2${}) at the n-hexane -- water interface.

To conclude, the analysis of scattering data has shown that, with decreasing temperature $T$, a two-dimensional condensation transition of the C$_{30}$-alcohol to the Gibbs liquid monolayer at the interface at the temperature $T_c$ is followed at the temperature $T^*$ by a transition to its multilayer adsorption. We believe that this adsorption is caused by an increase in the concentration of micelles in the 100- to 200-\AA-thick surface layer. The observation of such transitions in two-component adsorbed fluoroalkanol films and in C$_{30}$-alcohol and C$_{30}$-acid single-component films at the n-hexane–water interface was reported earlier [9,30,31].

The work at the National Synchrotron Light Source was supported by the US Department of
Energy (contract no. DE-AC02-98CH10886). The work at the X19C beamline was supported by the
ChemMatCARS National Synchrotron Resource, University of Chicago, University of Illinois at Chicago,
and Stony Brook University.

\end{document}